\documentclass[prd,aps,showkeys,superscriptaddress,two column]{revtex4}
\usepackage{amsmath}
\usepackage{amssymb}
\usepackage{amsthm}
\usepackage{mathrsfs}
\usepackage{graphicx}
\usepackage{fancyhdr}
\usepackage{array}
\usepackage{simplewick}
\usepackage{latexsym}
\usepackage[all]{xy}
\usepackage{eufrak}
\usepackage{euscript}
\usepackage{enumerate}
\usepackage{dsfont}
\usepackage{slashed}
\usepackage{hyperref}

\begin{document}

%%%%%%%%%%%%%%%%%%%%%%%%%%%%%%%%%%%%%%%%%%%%%%%%%%%%%%%%%%%%%

\title{Julia-Toulouse approach to $(d+1)$-dimensional bosonized Schwinger model with an application to large $N$ QCD}

\author{M. S. Guimaraes}
\email{msguimaraes@uerj.br}
\affiliation{Instituto de F\'\i sica, Universidade do Estado do Rio de Janeiro, 20550-013, Rio de Janeiro, Brazil}

\author{R. Rougemont}
\email{romulo@if.ufrj.br}
\affiliation{Instituto de F\'\i sica, Universidade Federal do Rio de Janeiro, 21941-972, Rio de Janeiro, Brazil}

\author{C. Wotzasek}
\email{clovis@if.ufrj.br}
\affiliation{Instituto de F\'\i sica, Universidade Federal do Rio de Janeiro, 21941-972, Rio de Janeiro, Brazil}

\author{C. A. D. Zarro}
\email{carlos.zarro@if.ufrj.br}
\affiliation{Instituto de F\'\i sica, Universidade Federal do Rio de Janeiro, 21941-972, Rio de Janeiro, Brazil}

%\date{\today}

\begin{abstract}
The Julia-Toulouse approach for condensation of charges and defects is used to show that the bosonized Schwinger model can be obtained through a condensation of electric charges in $1+1$ dimensions. The massive model is derived by taking into account the presence of vortices over the electric condensate, while the massless model is obtained when these vortices are absent. This construction is then straightforwardly generalized for arbitrary $d+1$ spacetime dimensions. The $d=3$ case corresponds to the large $N$ chiral dynamics of $SU(N)$ QCD in the limit $N\rightarrow\infty$.
\end{abstract}

%\pacs{Valid PACS appear here}
% PACS, the Physics and Astronomy Classification Scheme.
% Valid PACS numbers may be entered using the \verb+\pacs{#1} command.

% Use showkeys class option if keyword display desired

%%%%%%%%%%%%%%%%%%%%%%%%%%%%%%%%%%%%%%%%%%%%%%%%%%%%%%%%%%%%%

\maketitle

%\tableofcontents

\section{Introduction}
\label{sec:intro}

Schwinger model \cite{schwinger,various} is the name given to electrodynamics in $1+1$ dimensions. This model was originally examined by Schwinger as an example where mass and gauge invariance coexists compatibly. While the classical theory is confining due to the linear behavior of the Coulomb interaction in $1+1$ dimensions, quantum effects can modify this picture. The quantum theory with massless fermions is exactly solvable (i.e., all the Green's functions of the model can be obtained in closed form) and electric probe charges are screened due to the mass acquired by the gauge boson due to fermionic fluctuations \cite{schwinger,various,gross,abdalla}. On the other hand, in the quantum theory with massive fermions (which is not exactly solvable), electric probe charges interact via an effective potential that features both, a screening piece and a linear confining term. For large inter-charge separations the confining term prevails as long as the theta-vacuum angle is different from $\pi$ and the probe charges are not integer multiples of the dynamical fermionic charges, in which case the confining term vanishes \cite{various,gross,abdalla}.

The quantized versions of both, the massless and massive Schwinger models, possess exact bosonic representations in $1+1$ dimensions, where the fermions are replaced by scalar bosons. This quantum fermion-boson map, called \textit{bosonization}, has been obtained in the literature by using different techniques, as for instance, by computing the fermionic determinant, through loop and derivative expansions, etc (see \cite{stone} for a review on the subject).

In this Letter we present a new path for obtaining the bosonized versions of both, the massless and massive Schwinger models. This new construction is realized by considering a condensation of electric charges in $1+1$ dimensions via the Julia-Toulouse approach (JTA) for condensation of charges and defects \cite{jt,qt,artigao,mbf-m}. The massless Schwinger model is obtained when there are no vortices over the electric condensate (complete condensation) and the massive Schwinger model is obtained by taking into account the contribution of these defects (incomplete condensation).  

The 1-form $A_1$ is the gauge connection with maximal rank that one can define in $1+1$ dimensions. As observed in \cite{aurilia}, the general properties of the bosonized version of the Schwinger model are associated with this fact and, consequently, these general properties could be extended for arbitrary $d+1$ dimensions by working in terms of a maximal rank gauge connection $A_d$. By taking this observation into account, we show that the bosonized versions of the massless and massive Schwinger models can be straightforwardly generalized for arbitrary $d+1$ spacetime dimensions by considering the condensation of $d$-currents $J_d$ minimally coupled to a maximal rank gauge connection $A_d$. Such a $(d+1)$-dimensional generalization of the bosonized versions of the massless and massive Schwinger models, however, contrary to what happens in $1+1$ dimensions, is not associated to a fermion-boson map in higher dimensional spacetimes. Indeed, as we shall discuss, the $(3+1)$-dimensional extension of the bosonized version of the massive Schwinger model corresponds to the large $N$ chiral dynamics of the $(3+1)$-dimensional $SU(N)$ QCD in the limit $N\rightarrow\infty$, extending a previous observation reported in \cite{aurilia}.

For uses of the maximal rank gauge connection in other physical scenarios, see for example \cite{cc,misc}. The results that we shall present here may be of interest for the lines of investigation developed in these references.

The JTA \cite{jt,qt} is a prescription used to construct a low energy effective theory describing a system with condensed charges or defects, having previous knowledge of the model that describes the system in the regime with diluted charges or defects and also of the symmetries expected for the condensed regime. Based mainly on \cite{jt,qt}, and taking also into account the ideas developed in \cite{banks,mvf} regarding the formulation of ensembles of charges and defects, we introduced in \cite{artigao,mbf-m} a generalization of the JTA, which we shall use in this Letter. In particular, we are going to work with the dual JTA \cite{artigao}, which is defined in the dual picture to the one originally proposed in \cite{qt}.

%In \cite{qt}, the condensing charges are described in a picture where they couple non-minimally to a given $p$-form. In the case approached here, since we are going to deal with condensing electric charges, a non-minimal coupling of these charges would be made with the electromagnetic dual of the 1-form gauge field, which in $1+1$ dimensions would be a $-1$-form, i.e., a $p$-form with negative rank. Actually, this is a general feature of a maximal rank gauge connection in $d+1$ dimensions: its electromagnetic dual would be $-1$-form. To avoid the introduction of such a mathematically awkward object, the dual JTA is used, since in this picture the condensing topological currents $J_d$ are minimally coupled to the maximal rank gauge connection $A_d$.

\section{The bosonized Schwinger model as an electric condensate}
\label{sec:jta}

In this Letter we shall use natural units of $c=\hbar=1$. We begin working in Minkowski spacetime $\mathbb{R}^{1,1}$.

The partition function describing the interaction of a gauge boson with external electric charges dilutely distributed through the space is given by:
\begin{align}
Z_d[J_1]&=\int_{G.F.}\mathcal{D}A_1 \exp\left\{i\int_{\mathbb{R}^{1,1}}\left[-\frac{1}{2}dA_1\wedge*dA_1+\right.\right.\nonumber\\
&\left.\left.-eA_1\wedge *J_1\right]\right\},
\label{eq:1}
\end{align}
where $J_1=\delta\Sigma_2$ is the electric current that localizes the world-line of the electric charge $e$, the physical boundary of the world-surface of the electric Dirac string \cite{dirac} (electric Dirac brane) localized by the Chern-Kernel $\Sigma_2$. The acronym ``G.F.'' in the functional integral stands for some arbitrary gauge fixing procedure that must be implemented at some stage of the calculations.

By integrating out the gauge field we obtain a Coulomb interaction between the classical electric charges, which is confining in this dimensionality. This is easy to understand in physical terms: with only one spatial dimension, the electric field flows as it was in the interior of a straight flux tube and, consequently, the Coulomb potential is linear in the inter-charges separation in $1+1$ dimensions. However, this picture is changed when we take quantum effects into account. For this sake, we shall apply the JTA in the sequel to study the effects produced by a condensation of electric charges (after this step, one can include electric probe charges into the system and compare the new results with the classical confining picture).

The JTA has a very definite physical meaning: it constitutes a mass generation mechanism for arbitrary $p$-forms due to the condensation of topological $p$-currents (charges or defects). The form of the effective theory describing the low energy excitations of the condensate of topological currents is obtained via JTA by following the cornerstone of effective field theories, that is the symmetry content of a determined physical system. A given symmetry content strongly constrains the form of the effective model for the condensate of topological currents, such that one can reach a great variety of different physical phenomenologies depending on the nature of the condensing currents: some examples, including condensates with and without violation of discrete spacetime symmetries and also Lorentz invariance violation (LIV), can be found in reference \cite{artigao}. The specific way by which the JTA links the diluted and condensed phases is through the addition of a certain fugacity in the partition function, corresponding to the complex exponential of an activation term for the condensing currents, which is responsible for accounting for changes in the number of charges or defects in the system, allowing a proliferation of these objects. The specific form of the fugacity is determined by the symmetries expected for the condensed phase and by the requirement that the effective theory obtained describes the low-lying excitations of the established condensate: in this way, only terms in lowest order in a derivative expansion of the Chern-Kernels of the condensing currents compatible with a given set of symmetries are retained in the expression for the fugacity. Such an approach is sufficient to account for the dominant contribution for the dynamics of the condensate in the low energy regime. Furthermore, as a consequence of this condensation process, there is generation of mass for the initially massless $p$-forms present in the diluted phase.

To implement the JTA here, we begin by adding to the Boltzmann factor in (\ref{eq:1}) an activation term for the electric currents (which effectively gives dynamics to the electric Dirac branes) such that it preserves the relevant symmetries of the system ($P$, $T$, Lorentz and the local gauge symmetry) and gives the dominant contribution for the dynamics of the electric condensate in the low energy regime \cite{mvf,artigao,mbf-m}:
\begin{align}
S_{activation}[J_1]&=\int_{\mathbb{R}^{1,1}}\frac{-1}{2\Lambda}J_1\wedge *J_1\nonumber\\
&=\int_{\mathbb{R}^{1,1}}\frac{1}{2\Lambda}d*\Sigma_2\wedge *d*\Sigma_2,
\label{eq:2}
\end{align}
where $\Lambda$ is an adimensional free parameter of the JTA which we shall fix afterwards by comparison with the results obtained by using bosonization techniques. Introducing also a formal sum over (the branes Poincare-dual to) $*\Sigma_2$, we obtain the partition function defining the electric condensed regime:
\begin{align}
Z_c&:=\sum_{\left\{*\Sigma_2\right\}}\int_{G.F.}\mathcal{D}A_1  \exp\left\{i\int_{\mathbb{R}^{1,1}}\left[-\frac{1}{2}dA_1\right.\right.\wedge*dA_1+\nonumber\\
&\left.\left.-eA_1\wedge d*\Sigma_2+\frac{1}{2\Lambda}d*\Sigma_2\wedge *d*\Sigma_2\right]\right\}.
\label{eq:3}
\end{align}
Introducing into (\ref{eq:3}) the identity $\mathds{1}=\int\mathcal{D}*P_2\delta[*P_2 - *\Sigma_2]$, we rewrite the partition function for the electric condensed regime as:
\begin{align}
&Z_c=\int_{G.F.}\mathcal{D}A_1\mathcal{D}*P_2\left(\sum_{\left\{*\Sigma_2\right\}}\delta[*P_2 - *\Sigma_2]\right)
\nonumber\\
&\exp\left\{i\int_{\mathbb{R}^{1,1}}\left[-\frac{1}{2}dA_1\wedge*dA_1-eA_1\wedge d*P_2 +\right.
\right.\nonumber\\
&\left.\left. + \frac{1}{2\Lambda}d*P_2\wedge *d*P_2\right]\right\}.
\label{eq:4}
\end{align}
Next, we make use of a generalized version of the Poisson's identity (GPI) \cite{mvf,dafdc}:
\begin{align}
\sum_{\left\{*\Sigma_2\right\}}\delta[*P_2 - *\Sigma_2]&=\sum_{\left\{*\Omega_0\right\}}\exp\left\{2\pi i\int_{\mathbb{R}^{1,1}} *\Omega_0\wedge *P_2\right\}\nonumber\\
&=\sum_{\left\{\Omega_0\right\}}\exp\left\{2\pi i\int_{\mathbb{R}^{1,1}} d^2x\, \Omega_0 (*P_2)\right\},
\label{eq:5}
\end{align}
where $\Omega_0$ is the brane Poisson-dual to $\Sigma_2$. The GPI works as a geometric analogue of the Fourier transform: when the brane configurations on the left hand side of (\ref{eq:5}) proliferate (condense), the brane configurations on the right hand side become diluted and vice-versa (see appendix A of \cite{dafdc} for a detailed discussion and derivation of the GPI in the general case). Hence, the proliferation of the electric Dirac branes $\Sigma_2$ (which is directly associated to the proliferation of the electric currents $J_1$ that live on their boundaries) is accompanied by the dilution of the branes of complementary dimension $\Omega_0$ and vice-versa, what tells us that the branes $\Omega_0$ must be interpreted as vortices (defects) over the electric condensate.

Using (\ref{eq:5}) and redefining $*P_2=:\sqrt{\Lambda}\phi$, we rewrite (\ref{eq:4}) as:
\begin{align}
Z_c&=\int_{G.F.}\mathcal{D}A_1\mathcal{D}\phi
\exp\left\{i\int_{\mathbb{R}^{1,1}}\left[-\frac{1}{2}dA_1\wedge*dA_1 +\right.\right.\nonumber\\
&\left.\left.-e\sqrt{\Lambda}\,A_1\wedge d\phi + \frac{1}{2}d\phi\wedge *d\phi\right]\right\}Z_V[\phi],
\label{eq:6}
\end{align}
where:
\begin{align}
Z_V[\phi]=\sum_{\left\{\Omega_0\right\}}\exp\left\{i\int_{\mathbb{R}^{1,1}} d^2x\, 2\pi\sqrt{\Lambda}\,\Omega_0 \phi\right\},
\label{eq:7}
\end{align}
is the vortex partition function. Equations (\ref{eq:6}) and (\ref{eq:7}) constitute the result of the use of the JTA for deriving the effective low energy theory of an electric condensate in $1+1$ dimensions, being $\phi$ the scalar field describing the electric condensate.

Now, to make explicit contact with the Schwinger model, we must consider the vortex contribution formally encoded in (\ref{eq:7}).

\begin{itemize}
\item \underline{\textbf{Case 1: The massless Schwinger model}}
\end{itemize}

Let us suppose that the system is in a state with complete electric condensation, i.e., there are no vortices over the electric condensate such that the partition function (\ref{eq:7}) is trivial ($\Omega_0\rightarrow 0 \Rightarrow Z_V[\phi]\rightarrow\mathds{1}$). In this case, the partition function (\ref{eq:6}) gives the bosonized version of the massless Schwinger model:
\begin{align}
Z_c^{m=0}&=\int_{G.F.}\mathcal{D}A_1\mathcal{D}\phi
\exp\left\{i\int_{\mathbb{R}^{1,1}}\left[-\frac{1}{2}dA_1\wedge*dA_1 +\right.\right.\nonumber\\
&\left.\left.-\frac{e}{\sqrt{\pi}}\,A_1\wedge d\phi + \frac{1}{2}d\phi\wedge *d\phi\right]\right\},
\label{eq:8}
\end{align}
where we fixed the JTA parameter $\Lambda=\pi^{-1}$ by comparison with the result obtained using bosonization techniques (see, for example, \cite{gross}).

By including electric probe charges into this system via a minimal coupling with the gauge field, one finds that the probe charges are screened by the electric condensate (see, for example, \cite{abdalla} for a detailed discussion). This result already shows that the classical confining picture is changed when quantum effects (electric condensation) are taken into account.

\begin{itemize}
\item \underline{\textbf{Case 2: The massive Schwinger model}}
\end{itemize}

Let us suppose now that the system is in a state with incomplete electric condensation, i.e., there are vortices over the electric condensate such that the partition function (\ref{eq:7}) is not trivial. In the present case, one can give a precise prescription to realize the sum over point-vortices formally encoded in (\ref{eq:7}). For this sake, we are going to adopt the dilute gas approximation \cite{schakel,polyakov}.

We begin by Wick-rotating (\ref{eq:7}) to the euclidean space $\mathbb{R}^2$ ($t\mapsto -it_E, d^2x\mapsto -id^2x_E,\Omega_0\mapsto -i\Omega_0^E,\phi\mapsto \phi_E$), where the sum over vortices is translated into a sum over instantons \cite{schakel}:
\begin{align}
Z_V^E[\phi_E]=\sum_{\left\{\Omega_0^E\right\}}\exp\left\{-i\int_{\mathbb{R}^2} d^2x_E\, 2\sqrt{\pi}\,\Omega_0^E \phi_E\right\},
\label{eq:9}
\end{align}
where we used that $\Lambda=\pi^{-1}$. First we consider the contribution of a single instanton with winding number $+1$ for the partition function $Z_V^E[\phi_E]$:
\begin{align}
\int_{\mathbb{R}^2} d^2x^\alpha_E\, z&\exp\left\{-i\int_{\mathbb{R}^2} d^2x_E\,2\sqrt{\pi}\, \delta^{(2)}(x_E-x_E^\alpha)\phi_E(x_E)\right\}=\nonumber\\
&=\int_{\mathbb{R}^2} d^2x^\alpha_E\, z\exp\left\{-i2\sqrt{\pi}\,\phi_E(x^\alpha_E)\right\},
\label{eq:10}
\end{align}
where $z$ is the vortex fugacity (which gives the probability density of existence of a single instanton in space-time with winding number $+1$) and we are integrating over all the possible locations $x_E^\alpha$ of the instanton. Summing over all the possible configurations of the system with an arbitrary number of instantons with winding number $+1$, assuming that the instantons do not interact among themselves, we have for $Z_V^E[\phi_E]$:
\begin{align}
\sum_{N_+=0}^\infty&\frac{1}{N_+!}\left(\int_{\mathbb{R}^2} d^2x^\alpha_E\, z \exp\left\{-i2\sqrt{\pi}\,\phi_E(x^\alpha_E)\right\}\right)^{N_+}=\nonumber\\
&=\exp\left\{\int_{\mathbb{R}^2} d^2x_E\, ze^{-i2\sqrt{\pi}\,\phi_E(x_E)}\right\},
\label{eq:11}
\end{align}
where the factor $(N_+!)^{-1}$ was introduced due to the fact that the instantons are indistinguishable. Taking also into account the contribution of an arbitrary number of antinstantons with winding number $-1$ and neglecting the contribution of instantons and antinstantons with higher winding numbers (which are exponentially supressed in the partition function if we assume a small fugacity), we get (see also \cite{artigao}):
\begin{align}
Z_V^E[\phi_E]\approx&\exp\left\{\int_{\mathbb{R}^2} d^2x_E\, ze^{-i2\sqrt{\pi}\,\phi_E(x_E)}\right\}\times\nonumber\\
&\times\exp\left\{\int_{\mathbb{R}^2} d^2x_E\, ze^{+i2\sqrt{\pi}\,\phi_E(x_E)}\right\}=\nonumber\\
&=\exp\left\{\int_{\mathbb{R}^2} d^2x_E\, 2z\cos(2\sqrt{\pi}\,\phi_E(x_E))\right\}.
\label{eq:12}
\end{align}
Hence, the net effect of the instanton or vortex contribution is to generate a cossine for the scalar field \cite{schakel,polyakov}. Wick-rotating (\ref{eq:12}) back to Minkowski and substituting the result into (\ref{eq:6}), we get the bosonized version of the massive Schwinger model:
\begin{align}
Z_c^{m\neq 0}&=\int_{G.F.}\mathcal{D}A_1\mathcal{D}\phi
\exp\left\{i\int_{\mathbb{R}^{1,1}}\left[-\frac{1}{2}dA_1\wedge*dA_1 +\right.\right.\nonumber\\
&\left.\left.-\frac{e}{\sqrt{\pi}}\,A_1\wedge d\phi + \frac{1}{2}d\phi\wedge *d\phi+2z\cos(2\sqrt{\pi}\,\phi)d^2x\right]\right\}.
\label{eq:13}
\end{align}
If we compare (\ref{eq:13}) with the result obtained using bosonization techniques, we fix the vortex fugacity to be $z=\frac{me\exp(\gamma)}{4\pi^{3/2}}$, where $m$ is the fermion mass and $\gamma$ is the Euler constant (see, for example, \cite{gross}). We then realize that the condition for small fugacity (which allows one to ignore the contribution of vortices with higher winding numbers) corresponds to the small coupling regime.

By including electric probe charges into this system via a minimal coupling with the gauge field, one finds that the probe charges interact via an approximate effective potential that features two parts: a screening piece plus a linear confining term. The confining term prevails for large inter-charge separations, as long as the theta-vacuum angle is different from $\pi$ and the probe charges are not integer multiples of the charge of the electric condensate, in which case the confining term vanishes, restoring the screening phase. The theta-vaccum angle is introduced in the calculations as an integration constant in the evaluation of the interaction potential between the probe charges and corresponds to a (generally) non-vanishing background electric field (see, for example, \cite{abdalla} for a detailed discussion).

The connection between the vortex contribution and the mass of the fermions in $1+1$ dimensions is detailed discussed in \cite{schakel}. There, it is pointed out that an index theorem establishes the equality between the index of the massless Dirac operator (that is given by the difference between zero modes of the Dirac operator with positive and negative chiralities) and the Pontryagin index (topological charge or winding number) of the vortices. In the presence of vortices, this topological charge would be non-vanishing and, hence, there would be necessarily null eigenvalues of the massless Dirac operator in the fermionic determinant, in which case it would vanish. Therefore, for the massless Schwinger model, the vortex contribution is completely suppressed by the massless fermions. The situation is quite different in the massive case, since the massive Dirac operator has no zero modes and the vortex contribution is non-trivial in this case.

\section{$(d+1)$-dimensional generalization of the bosonized Schwinger model via JTA}
\label{sec:general}

One of the great advantages of the JTA is that it allows a straightforward generalization of the preceding construction for arbitrary $d+1$ spacetime dimensions. For this sake, we begin by considering the partition function describing the interaction of a maximal rank gauge connection $A_d$ with external topological currents $J_d$ dilutely distributed through a $(d+1)$-dimensional Minkowski spacetime with metric $\textrm{diag}(-,+,\cdots,+)$ \footnote{This is a particular case of the $p$-form electrodynamics for extended objects discussed in: C. Teitelboim, Phys. Lett. B \textbf{167}, 63 (1986).}:
\begin{align}
Z_d[J_d]&=\int_{G.F.}\mathcal{D}A_d \exp\left\{i\int_{\mathbb{R}^{1,d}}\left[\frac{(-1)^d}{2}dA_d\wedge*dA_d+\right.\right.\nonumber\\
&\left.\left.-eA_d\wedge *J_d\right]\right\},
\label{eq:14}
\end{align}
where $J_d=\delta\Sigma_{d+1}$. The JTA is implemented, as before, by adding to the Boltzmann factor in (\ref{eq:14}) the following activation term for the topological condensing $d$-currents:
\begin{align}
S_{activation}[J_d]&=\int_{\mathbb{R}^{1,d}}\frac{-1}{2\lambda^{1-d}}J_d\wedge *J_d\nonumber\\
&=\int_{\mathbb{R}^{1,d}}\frac{1}{2\lambda^{1-d}}d*\Sigma_{d+1}\wedge *d*\Sigma_{d+1},
\label{eq:15}
\end{align}
where $\lambda$ is a phenomenological JTA parameter with mass dimension, being the partition function for the condensed regime given by:
\begin{align}
Z_c&:=\sum_{\left\{*\Sigma_{d+1}\right\}}\int_{G.F.}\mathcal{D}A_d  \exp\left\{i\int_{\mathbb{R}^{1,d}}\left[\frac{(-1)^d}{2}dA_d\right.\right.\wedge*dA_d+\nonumber\\
&\left.\left.+(-1)^d eA_d\wedge d*\Sigma_{d+1}+\frac{1}{2\lambda^{1-d}}d*\Sigma_{d+1}\wedge *d*\Sigma_{d+1}\right]\right\}.
\label{eq:16}
\end{align}
By repeating the same steps between equations (\ref{eq:3}) and (\ref{eq:6}), we rewrite (\ref{eq:16}) as:
\begin{align}
Z_c&=\int_{G.F.}\mathcal{D}A_d\mathcal{D}\phi
\exp\left\{i\int_{\mathbb{R}^{1,d}}\left[\frac{(-1)^d}{2}dA_d\wedge*dA_d +\right.\right.\nonumber\\
&\left.\left.+(-1)^d mA_d\wedge d\phi + \frac{1}{2}d\phi\wedge *d\phi\right]\right\}Z_V[\phi],
\label{eq:17}
\end{align}
where $m:=e\lambda^{(1-d)/2}$ is the topological mass \footnote{The mass dimensions involved are: $[A_d]=[\phi]=(d-1)/2$, $[e]=(d+1)/2$, $[J_d]=1$ and $[\Sigma_{d+1}]=0$.} generated by the condensation of topological $d$-currents and:
\begin{align}
Z_V[\phi]=\sum_{\left\{\Omega_0\right\}}\exp\left\{i\int_{\mathbb{R}^{1,d}} d^{d+1}x\, 2\pi\lambda^{(1-d)/2}\,\Omega_0 \phi\right\},
\label{eq:18}
\end{align}
is the vortex partition function. As before, we can evaluate the vortex contribution approximately by considering a small vortex fugacity $z$ and the $(d+1)$-dimensional generalization of the bosonized version of the massive Schwinger model reads:
\begin{align}
Z_c=\int_{G.F.}\mathcal{D}A_d\mathcal{D}\phi
&\exp\left\{i\int_{\mathbb{R}^{1,d}}\left[\frac{(-1)^d}{2}dA_d\wedge*dA_d +\right.\right.\nonumber\\
&+(-1)^d mA_d\wedge d\phi + \frac{1}{2}d\phi\wedge *d\phi+\nonumber\\
&\left.\left.+2z\cos(2\pi\lambda^{(1-d)/2}\phi)d^{d+1}x\right]\right\}.
\label{eq:19}
\end{align}
The $(d+1)$-dimensional generalization of the bosonized version of the massless Schwinger model, corresponding to a complete condensation of $d$-currents, is recovered from (\ref{eq:19}) by taking the vortex fugacity $z$ to vanish.

\section{Application: the $d=3$ case and large $N$ chiral dynamics}
\label{sec:largeN}

Let us now consider the $d=3$ case of the preceding construction and its connection with the large $N$ chiral dynamics of the $(3+1)$-dimensional $SU(N)$ QCD \cite{witten-largeN}.

By writing down the vacuum-to-vacuum transition amplitude in a given theta-vacuum for QCD, one identifies a $CP$ violating term given by \cite{PQ}:
\begin{align}
\mathcal{L}_\theta=\theta\frac{g^2}{32\pi^2}\mbox{Tr}[\mathds{F}_2\wedge\mathds{F}_2],
\label{eq:20}
\end{align}
where $g$ is the QCD coupling constant, $\mathds{F}_2=d\mathds{A}_1+\mathds{A}_1\wedge\mathds{A}_1$ and $\mathds{A}_1=A_1^a\mathds{T}^a$, with $\left\{\mathds{T}^a\right\}$ being the set of the $N^2-1$ generators of the $\mathfrak{su}(N)$ algebra. The action associated to the Lagrangian density (\ref{eq:20}) is a surface term, which is non-vanishing due to instanton configurations of the non-Abelian connection.

It was pointed out in \cite{luscher} that one can rewrite the theta-term (\ref{eq:20}) in terms of an Abelian 3-form $A_3$ according to:
\begin{align}
\mathcal{L}_\theta=\theta dA_3=-\theta *F,
\label{eq:21}
\end{align}
where $F:=*F_4=*dA_3$, being the Abelian 3-form $A_3$ a composite field defined by the trace of the non-Abelian Chern-Simons 3-form. It can be shown that under an arbitrary $SU(N)$ gauge transformation, $A_3$ transforms like an Abelian connection; furthermore, it can be shown that the 2-point correlation function of the composite field $A_3$ corresponds to a Coulomb propagator, and hence $A_3$ behaves as a massless colorless collective field propagating a long-range interaction \cite{luscher,gabadadze-largeN}. At this point, one could aim to construct an effective action for the Abelian field $A_3$. The simplest one corresponds to the very low energy limit (where the masses of the quarks and hence, the masses of the mesons, are taken to infinity) of the very large $N$ approximation for the effective action of QCD describing the pseudo-Goldstone bosons (pseudoscalar mesons) associated to the chiral symmetry breaking produced by the quark condensate \cite{gabadadze-largeN}:
\begin{align}
S^{eff}_d[A_3,J_3]\sim\int_{\mathbb{R}^{1,3}}\left[-\frac{1}{2\chi}dA_3\wedge*dA_3+\theta dA_3-eA_3\wedge *J_3\right],
\label{eq:22}
\end{align}
where $\chi$ is the vacuum topological susceptibility and where the mesons were integrated out in the above referred approximations and we also added the last term in (\ref{eq:22}) corresponding to a source $J_3$ for the Abelian field $A_3$. Rescaling $A_3\mapsto \sqrt{\chi}A_3$ and absorbing a constant factor of $\sqrt{\chi}$ into the parameters $e$ and $\theta$, we rewrite (\ref{eq:22}) as:
\begin{align}
S^{eff}_d[A_3,J_3]\sim\int_{\mathbb{R}^{1,3}}\left[-\frac{1}{2}dA_3\wedge*dA_3+\theta dA_3-eA_3\wedge *J_3\right],
\label{eq:23}
\end{align}
which, apart from the theta-term, is exactly the diluted phase for the sources $J_3$ in the $d=3$ case of (\ref{eq:14}).

It is important to notice that the field strength $F=*F_4=*dA_3$ propagates no dynamical degrees of freedom in $3+1$ dimensions, since it can be shown from the equations of motion coming from (\ref{eq:23}), that $F$ is just a constant in spacetime \cite{gabadadze-largeN}. Hence, the composite field $A_3$, although responsible for setting a constant background field $F$ into the theory, does not imply in the presence of any dynamical massless colorless collective excitations of the gluons in the QCD spectrum, what is desirable, otherwise it would enter in conflict with the expectation that there is a mass gap in QCD \cite{gabadadze-largeN}.

If the sources $J_3$ undergo an incomplete condensation process, there being vortices over the bubble condensate, the effective action for the condensed phase can be read off from (\ref{eq:19}), taking into account the theta-term present in (\ref{eq:23}):
\begin{align}
S^{eff}_c[A_3,\phi]&=\int_{\mathbb{R}^{1,3}}\left[-\frac{1}{2}dA_3\wedge*dA_3+\theta dA_3-mA_3\wedge d\phi+\right.\nonumber\\ &\left.+\frac{1}{2}d\phi\wedge*d\phi+2z\cos\left(\frac{2\pi}{\lambda}\phi\right) d^4x\right].
\label{eq:24}
\end{align}
Rescaling $A_3\mapsto\frac{\lambda}{2\pi}A_3$ and $\phi\mapsto\frac{\lambda}{2\pi}\phi$, we get:
\begin{align}
S^{eff}_c[A_3,\phi]&=\int_{\mathbb{R}^{1,3}}\left[\frac{\lambda^2}{4\pi^2}\left(-\frac{1}{2}dA_3\wedge*dA_3+\frac{2\pi\theta}{\lambda} dA_3+\right.\right.\nonumber\\
&\left.\left.-\frac{e}{\lambda}A_3\wedge d\phi+\frac{1}{2}d\phi\wedge*d\phi\right)+2z\cos(\phi) d^4x\right].
\label{eq:25}
\end{align}
For (\ref{eq:25}) to make sense in the context of an $1/N$ expansion, we identify \footnote{As discussed in pages 252 and 253 of \cite{smilga}, in this context, all the coefficients of the action must scale with $N$.} $\lambda^2=4\pi^2\bar{\lambda}^2N$, with $\bar{\lambda}^2$ a constant with mass dimension 2, and $2z=\bar{\lambda}^2N\bar{z}$, with $\bar{z}$ a constant with mass dimension 2, such that we rewrite (\ref{eq:25}) as:
\begin{align}
S^{eff}_c[A_3,\phi]&=N\bar{\lambda}^2\int_{\mathbb{R}^{1,3}}\left[-\frac{1}{2}dA_3\wedge*dA_3+\frac{\theta}{\bar{\lambda}\sqrt{N}} dA_3+\right.\nonumber\\
&\left.-\frac{e}{2\pi\bar{\lambda}\sqrt{N}}A_3\wedge d\phi+\frac{1}{2}d\phi\wedge*d\phi+\bar{z}\cos(\phi) d^4x\right],
\label{eq:26}
\end{align}
from which follows that, in the limit $N\rightarrow\infty$ in the partition function associated to (\ref{eq:26}), the relevant field configurations are the classical configurations extremizing the action.

Integrating out $A_3$, we obtain:
\begin{align}
S^{eff}_c[\phi]=N\bar{\lambda}^2\int_{\mathbb{R}^{1,3}}\left[\frac{1}{2}d\phi\wedge*d\phi-V(\phi)d^4x\right],
\label{eq:27}
\end{align}
where the potential energy is given by:
\begin{align}
V(\phi)&=\frac{1}{2}\left(\frac{e}{2\pi\bar{\lambda}\sqrt{N}}\phi-\frac{\theta}{\bar{\lambda}\sqrt{N}}\right)^2-\bar{z}\cos(\phi) \nonumber\\
&=\frac{e^2}{8\pi^2\bar{\lambda}^2N}(\phi-\theta)^2-\bar{z}\cos(\phi),
\label{eq:28}
\end{align}
where in the passage to the second line we absorbed a constant factor of $\frac{2\pi}{e}$ into the arbitrary constant $\theta$-parameter. From (\ref{eq:28}) we see that the mass of the pseudoscalar boson $\phi$ is given by:
\begin{align}
M_\phi^2=\frac{e^2}{4\pi^2\bar{\lambda}^2N}+\bar{z},
\label{eq:29}
\end{align}
which decreases as $1/N$ tending to a constant in the limit $N\rightarrow\infty$. But this is exactly the behaviour of the mass of the $\eta '$ meson in the large $N$ approximation \cite{smilga}! In fact, the potential energy (\ref{eq:28}) has exactly the same form of the potential energy obtained in \cite{witten-largeN} by considering the large $N$ effective chiral action for QCD with the quark masses fixed in the limit $N\rightarrow\infty$, case in which the different flavors are decoupled and the potential energy (\ref{eq:28}) refers to a given flavor. Hence, we can make the following identification of parameters by comparison with \cite{witten-largeN}: $N\bar{\lambda}^2=F_\pi^2$, $\frac{e^2}{4\pi^2\bar{\lambda}^2}=a$ and $\bar{z}=\mu^2$, where $F_\pi$ is the pion decay constant, $\mu$ is a constant coming from the quark mass matrix and $a$ is a constant of order 1 when $N\rightarrow\infty$.

As observed in \cite{witten-largeN}, the potential energy (\ref{eq:28}), obtained in \cite{witten-largeN} from the effective chiral action of $SU(N)$ QCD in $d=3$ with the quark masses fixed in the limit $N\rightarrow\infty$, is the same one obtained by bosonizing the massive quantum electrodynamics in $d=1$, two different and apparently uncorrelated theories. This fact is made clear in the formalism developed here, given that both models can be seen as particular cases, with $d=3$ and $d=1$, respectively, of the effective theory (\ref{eq:19}) describing a condensate of topological currents of maximal rank. Notice also that the generalized version of the massless Schwinger model in $d=3$ was already reported in \cite{aurilia} to be connected with the large $N$ chiral dynamics of $SU(N)$ QCD. As commented below, this is the case for a massless quark. Here we extended this connection by considering the generalized version of the massive Schwinger model in $d=3$, which corresponds to the realistic case with massive quarks.

Let us put some results into perspective. As stated before, the effective theory (\ref{eq:23}) is the limit of the effective action (\ref{eq:26}) when the quark mass goes to infinity. Given the above identification of parameters, this means that the vortex fugacity goes to infinity, such that, in order to keep the action of the system finite, $\phi$ must tend to a constant corresponding to its VEV in this limit, $2\pi n,\,n\in\mathbb{Z}$ \cite{witten-largeN}, and one recovers from (\ref{eq:26}) the action for the diluted phase (\ref{eq:23}) \footnote{Notice that for a constant $\phi=2\pi n,\,n\in\mathbb{Z}$ in the limit $\mu\rightarrow\infty\Rightarrow \bar{z}\rightarrow\infty$, the cosine term in (\ref{eq:26}) is a divergent constant, which is physically irrelevant, and $d\phi=0$.}: from the JTA point of view, this is compatible with the fact that for a huge vortex fugacity, there are many vortices over the system, what implies in the destruction of the bubble condensate. Another interesting case is seen in the opposite limit of zero quark mass (chiral limit), when we have $\mu=0\Rightarrow \bar{z}=0$: from the JTA point of view, this means that there are no vortices over the system (complete bubble condensation) and, in this case, the $\theta$-parameter can be eliminated from the theory via field redefinitions, such that there is no more $CP$ violation.

\section{Conclusion}
\label{sec:conclusion}

In this Letter we showed how the bosonized versions of the massless and massive Schwinger models can be constructed via JTA by considering a condensation of electric charges in $1+1$ dimensions. The massless case is obtained when there are no vortices over the electric condensate (complete condensation) and the massive case is derived when these defects are present (incomplete condensation). We then discussed the $(d+1)$-dimensional generalization of both, the massless and massive bosonized Schwinger models, associating their emergence with the condensation of topological $d$-currents minimally coupled to a maximal rank gauge connection $A_d$. In $d=3$, this generalization gives the large $N$ approximation for the effective action of $SU(N)$ QCD in the limit $N\rightarrow\infty$ \cite{witten-largeN}.  

As a final remark, we point out the fact that the electric condensate interpretation obtained here via JTA for fermionic radiative corrections in the 2-dimensional electrodynamics is analogous to the induction of the Chern-Simons term via JTA due to a $P$ and $T$ violating electric condensate in 3-dimensional electrodynamics \cite{santiago,artigao}: the Maxwell-Chern-Simons theory \cite{mcs1,mcs2,mcs3} constitutes the low energy effective theory derived by integrating out the fermionic degrees of freedom in $2+1$ dimensions. A mass term for the fermions in $2+1$ dimensions violates $P$ and $T$ and even for massless fermions these discrete space-time symmetries are violated due to the parity anomaly. This is the reason for electric condensates with different symmetries in $1+1$ and $2+1$ dimensions. These two explicit examples of fermionic radiative corrections being described as electric condensation processes show that the JTA can be made more general than just a dual description of the Higgs mechanism as originally proposed in \cite{qt}. This conclusion is rather reinforced by the fact that there is no Higgs mechanism in $1+1$ dimensions \cite{coleman}, but still the Schwinger mechanism can be described via JTA as showed in this Letter. Also, the Higgs mechanism is a mass generation mechanism for 1-form gauge fields, while the mass generation mechanism associated to the condensation of arbitrary extended topological $p$-currents described by the JTA can give mass for arbitrary $p$-form gauge fields.

\section{Acknowledgements}

We thank Conselho Nacional de Desenvolvimento Cient\'ifico e Tecnol\'ogico (CNPq) for financial support.

\end{document}